\documentclass[english,aps,prb,superscriptaddress,
amsmath,amssymb,preprintnumbers,twocolumn,floatfix,showpacs,showkeys,10pt]{revtex4-1}
\pdfoutput=1
\usepackage{filecontents}
\usepackage[T1]{fontenc}
\usepackage[utf8]{inputenc}
\usepackage[a4paper]{geometry}
\usepackage{color}
\usepackage[english]{babel}
\usepackage{float}
\usepackage{mathrsfs}
\usepackage{amsmath}
\usepackage{amssymb}
\usepackage{graphicx}
\usepackage{esint}
\usepackage[unicode=true,pdfusetitle,
 bookmarks=true,bookmarksnumbered=false,bookmarksopen=false,
 breaklinks=false,pdfborder={0 0 0},backref=false,colorlinks=true]
 {hyperref}
\hypersetup{
 linkcolor=blue,citecolor=blue,urlcolor=blue}

\usepackage{newfloat}
\DeclareFloatingEnvironment[name={Supplementary Figure}]{suppfigure}

\makeatletter

\providecommand{\tabularnewline}{\\}

 \@ifundefined{textcolor}{}
 {%
   \definecolor{BLACK}{gray}{0}
   \definecolor{WHITE}{gray}{1}
   \definecolor{RED}{rgb}{1,0,0}
   \definecolor{GREEN}{rgb}{0,1,0}
   \definecolor{BLUE}{rgb}{0,0,1}
   \definecolor{CYAN}{cmyk}{1,0,0,0}
   \definecolor{MAGENTA}{cmyk}{0,1,0,0}
   \definecolor{YELLOW}{cmyk}{0,0,1,0}
 }

\usepackage{color}

\usepackage{cancel} 
\usepackage{slashed}

\renewcommand{\emph}[1]{\textit{#1}}

\begin{document}

\title{Time-dependent Simulation and Analytical Modelling of Electronic
Mach-Zehnder Interferometry with Edge-states wave packets}
\author{Andrea Beggi}
\email[Corresponding author: ]{andrea.beggi@unimore.it}
\affiliation{Dipartimento di Scienze Fisiche, Informatiche e Matematiche, Universit{\`a} degli Studi 
di Modena e Reggio Emilia, Via Campi 213/A, 41125 Modena, Italy}
\author{Paolo Bordone}
\affiliation{Dipartimento di Scienze Fisiche, Informatiche e Matematiche, Universit{\`a} degli Studi 
di Modena e Reggio Emilia, Via Campi 213/A, 41125 Modena, Italy}
\affiliation{S3, Istituto Nanoscienze-CNR, Via Campi 213/A, 41125 Modena, Italy}
\author{Fabrizio Buscemi}
\affiliation{Dipartimento di Scienze Fisiche, Informatiche e Matematiche, Universit{\`a} degli Studi 
di Modena e Reggio Emilia, Via Campi 213/A, 41125 Modena, Italy}
\author{Andrea Bertoni}
\email[Corresponding author: ]{andrea.bertoni@unimore.it}
\affiliation{S3, Istituto Nanoscienze-CNR, Via Campi 213/A, 41125 Modena, Italy}

\begin{abstract}
We compute the exact single-particle time-resolved dynamics of electronic Mach-Zehnder interferometers based on Landau edge-states transport, and assess the effect of the spatial localization of carriers on the interference pattern.
The exact carrier dynamics is obtained by solving numerically the time-dependent Schr\"odinger equation with a suitable 2D potential profile reproducing the interferometer design. 
An external magnetic field, driving the system to the quantum Hall regime with filling factor one, is included.
The injected carriers are represented by a superposition of edge states and their interference pattern reproduces the results of [Y. Ji \emph{et al.}, \href{http://www.dx.doi.org/10.1038/nature01503}{Nature {\bf 422}, 415 (2003)}].
By tuning the system towards different regimes, we find two additional features in the transmission spectra, both related to carrier localization, namely a damping of the Aharonov–Bohm oscillations with increasing difference in the arms length, and an increased mean transmission that we trace to the energy-dependent transmittance of quantum point contacts.
Finally, we present an analytical model, also accounting for the finite spatial dispersion of the carriers, able to reproduce the above effects.
\end{abstract}

\pacs{73.43.-f, 85.35.Ds, 73.23.-b}
\keywords{Edge states; Time-dependent simulation; Mach-Zehnder interferometer; Aharonov-Bohm oscillations.}

\maketitle

\section{Introduction}


Edge states (ESs) are chiral one-dimensional conductive
channels which arise in a 2D electron gas (2DEG) in the Integer Quantum
Hall regime (IQHE)\cite{haug1993edgereport,Ferry09_Transport_Nano}.
They are essentially immune to backscattering and are characterized by very
long coherence lengths\cite{roulleau2008edge_coh_lenght}. 
Besides their remarkable interest for basic solid-state physics
and coherent electronic devices,
they are ideal candidates for both the so-called
``electron quantum optics''\cite{grenier2011electron} 
and ``flying qubit'' implementations of
quantum computing architectures\cite{benenti2004principles,Bertoni_Encycl_CSS}.

Indeed, several attempts to demonstrate single-qubit gate operations
and electronic interference in coupled quantum wires
were hampered by scattering and decoherence
processes\cite{ramamoorthy2007using,yamamoto2012electrical,bautze2014theoretical}.
On the other hand, experimental
realizations of electronic interferometry 
based on edge-channel transport seem more mature, to the point of
demonstrating not only single-electron\cite{JiMZI03,PhysRevLett.108.196803_Bocquillon}
but also two-electron interference\cite{Neder_DoubleMZI}.
Also, ``which-path'' detectors or quantum
erasers\cite{Neder_PRL_detectoreraser,Kang_MZIeraser,Dressel_which_path,BuscemiCoupledMZI,Weisz_ERASER}
have been implemented and the formation of quantum
entanglement between indistinguishable particles has been
demonstrated\cite{Neder_DoubleMZI,Samuelsson_HBT,Bocquillon_2Indisting_elec,PhysRevLett.101.166802_Olkhovskaya,PhysRevB.91.115438_Rossello}.

While specific features of electronic Mach-Zehnder Interferometers (MZIs)\cite{Neder_PRL_MZILobes}
are still under investigation and involve possibly many-particle
effects\cite{Dinh13Analytical_MZI_model,Rufino13Analytical_MZI_mode,PhysRevB.91.115438_Rossello},
their basic functioning can be explained in terms of stationary electron waves
interference, like their more common optical counterpart.
However, contrary to photons, the time that a conduction electron
takes to cross the interferometer can be comparable to switching times
of typical microelectronic devices, and its spatial localization
can be much less than the dimensions of a single arm of the interferometer.
Thus, understanding the detailed quantum dynamics of the carriers,
beyond the simple plane wave model is critical for the design of
novel devices based on ES transport.
This is even more relevant if the device is intended to process the
information encoded in a single particle at a time, a task which requires
a high resolution both in time and space. 

In fact, the injection of single-electron excitations that propagate along
ESs has been recently demonstrated\cite{Feve25052007,PhysRevLett.108.196803_Bocquillon,PhysRevLett.111.216807_Fletcher}:
with a proper time modulation of the injecting pulse,
they consists of Lorentzian-shaped excitations above the Fermi sea,
termed \emph{levitons}\cite{JMathPhys37.4845_Levitov,PhysRevLett.97.116403_Keeling,Nature502.659_Dubois}.  However, leviton-based MZI has not been realized experimentally so far.
The majority of the literature on electron MZIs in ESs deals with electronic
currents\cite{JiMZI03,Neder_PRL_MZILobes,Neder_PRL_detectoreraser,Neder_DoubleMZI,Weisz_ERASER},
and the prevalent theoretical models for the transport are based upon delocalized
scattering states\cite{Samuelsson_HBT,Kang_MZIeraser,Dressel_which_path,PhysRevB.77.115119_Kazymyrenko,PhysRevB.72.125320_Chung,BuscemiCoupledMZI},
with few notable exceptions\cite{PhysRevB.84.081303_Haack,NatCommun.5.3844_Gaury}.

In this work we address, both from numerical
and analytical points of view, the interference
properties of a Mach-Zehnder device 
based on ES channels tuned to filling factor $\nu$=1 and quantum point
contacts used as splitting elements,
when the electron travelling inside it is strongly localized.
We first use a numerical approach, which, unlike other recent
works\cite{kotimaki2012time,kotimaki2013TD-DFT},
is based upon the direct solution of the effective-mass time-dependent
Schrödinger equation, in presence of an external magnetic field.
The electrons travelling inside the device are localized wave packets (WPs)
of ESs, and are propagated by a generalized split-step Fourier 
method\cite{Weideman_SplitStepFourier,Taha_SplitStep,kramer2008efficient}.
The effects of the WP size on the transport process is analyzed.
Then, we develop an analytical model also accounting for the finite
spatial dispersion of the carriers.
The transmission coefficient of the device subject to Aharonov-Bohm (AB)
oscillations\cite{AharnovBohmPRL115} is obtained as a function of
the magnetic field and of the geometrical parameters
of the interfering paths, and the results are compared with the numerical
simulations.
Specifically, in Section \ref{sec:II-The-Physical-System},
we describe the initial electronic wave function and the
physical device used in our simulations.
In Section \ref{sec:IIINumerical-simulations}
we give some details on the numerical algorithm that we adopt
and we describe the results of the numerical simulations.
Section \ref{sec:IVTheoretical-Model-for}
is devoted to the development of an analytical model for the transport,
which takes into account the energy dispersion of the wave packet.
A comparison between the predictions of the latter model and the numerical
simulations is then presented in Section \ref{sec:VDiscussion}.
Finally, the conclusions are drawn in Section \ref{sec:VConclusions}.

\section{Physical system\label{sec:II-The-Physical-System}}

We consider a conduction-band electron in a 2DEG on the $x$-$y$ plane,
with charge $-e$ and effective mass $m^*$.
A uniform magnetic field $\vec{B}=(0,0,B)$
is applied along the $z$ direction
and a non-uniform electric field induces a local potential energy
$V(x,y)$.
The latter will represent the field generated by a polarized metallic gate
pattern above the 2DEG that will define energetically forbidden regions.
The generic Hamiltonian
$\hat{H}=(-i\hbar\vec{\nabla}+e\vec{A})^{2}/(2m^*)+V$
can be rewritten in a more explicit form
by substituting the magnetic vector potential $\vec{A}$
with its expression in the \emph{Landau gauge}
$\vec{A}=(0,\, Bx,\,0)$.
The 2D Hamiltonian for the electron in the 2DEG then reads
\begin{eqnarray}
\hat{H}=-\frac{\hbar^{2}}{2m^*}\frac{\partial^{2}}{\partial x^{2}}
-\frac{\hbar^{2}}{2m^*}\frac{\partial^{2}}{\partial y^{2}}-i\hbar
\frac{e B x}{m^* }\frac{\partial}{\partial y} \nonumber \\ 
+\frac{e^2 B^2}{2m^*}x^2+V(x,y).
\label{eq:2D_Hamilt}
\end{eqnarray}
We adopt a single-particle approach, thus neglecting the interaction with
other free electrons of the device.  
Our time-dependent numerical simulations and analytical model are based on
Eq.~(\ref{eq:2D_Hamilt}), as we explain in the
following. Before that, we must recall briefly the derivation of 
Landau states and of corresponding ESs\cite{jacoboni2010theory}.

Let us consider a region where the electric potential is invariant along $y$,
i.e. $V(x,y)=V(x)$. In the Landau gauge, the Hamiltonian (\ref{eq:2D_Hamilt})
shows an explicit translational symmetry along the $y$ direction,
and the quantum evolution of the particle can be factorized along the two axes.
In this case, the eigenstates of $\hat{H}$ have the form 
$\Psi(x,y)=\varphi(x)e^{iky}.$
In fact, the $y$-dependent part of the wave function (WF) is a plane
wave, while the $x$-dependent part (which depends also on the wavevector
$k$) is a solution of
$\hat{H}_{L}^{eff}\varphi_{n,k}(x)=E_{n}(k)\varphi_{n,k}(x)$, with the
following 1D effective Hamiltonian
\begin{equation}
\hat{H}_{L}^{eff}=\frac{-\hbar^{2}}{2m^*}\frac{\partial^{2}}{\partial x^{2}}
+\frac{1}{2}m^*\omega_{c}^{2}\left(x-x_{0}\right)^{2}+V(x),
\label{eq:Eff_Hamilt}
\end{equation}
where
\begin{equation}
  x_{0}(k) = -\frac{\hbar k}{eB},
\label{eq:k0-x0}
\end{equation}
and where $\omega_{c} = \frac{-eB}{m^*}$ is the cyclotron frequency.
Note that the parameter $x_{0}$ representing the center of the effective
parabolic confinement along $x$ depends on $k$, i.e. the wavevector in the
$y$ direction.
The discrete energy levels $E_{n}(k)$ associated with $\hat{H}_{L}^{eff}$
are the so-called \emph{Landau levels} (LLs).

If $V(x)=0$, the system eigenfunctions are the well-known localised
\emph{Landau states}.
However, if $V(x)$ is a step-like function, defining a sub-region
in which the electrons are confined, the Landau states with $x_0$
close to the edge have higher energy, and show a finite dispersion in $k$.
They become \emph{edge states},
which are delocalized WFs associated with a net probability
density flux, and which can act as 1D conductive channels.

Since we want to model a carrier as a propagating wave packet,
beyond the delocalized scattering state model,
we need to construct such WF as a proper combination
of ESs, rather than considering a single ES.
This is similar to representing a free electron by a 
minimum uncertainty wave packet rather than a plane wave,
thus including a finite uncertainty in both its position and
its kinetic energy.
However, in our case the magnetic field couples the two directions
and some care has to be taken, since the position along $x$ and the
velocity along $y$ are not independent.

We suppose that the injected electron lies in the first LL
and we represent it as a linear combination of ESs
with $n=1$, with Gaussian weights.
We choose a Gaussian shape for our initial WP since it represents a quantum
particle with the minimum quantum uncertainty in both its position and
momentum, and because its longitudinal spreading with the time is limited.
This allows us to follow clearly the dynamics of the localized WP without
substantial high-energy components forerunning the center of the WP and
leading to numerical instabilities. This choice also allows the derivation of
the approximated analytical model presented in Sect.~\ref{sec:IVTheoretical-Model-for}.
However, the qualitative results of our simulations do not depend on the
specific shape of the initial WP.
In fact, although we address the experimental regime of Ref.~\onlinecite{JiMZI03}
and not the propagation of levitons, as in Ref.~\onlinecite{Feve25052007},
we tested different shapes for the initial electron wave function: due to
the superposition principle the final state will be the composition of
the different scattering states taken with their initial weight.
\footnote{
In the Supplementary data, we compare the action of a QPC on a Gaussian
and a Lorentzian WP: in the second case the dynamics is more noisy and
the spread larger, but the effect of the QPC energy dependence is still evident.
}
Specifically, the wave function at the initial time will be
\begin{eqnarray}
\Psi_0(x,y) & = & \int dk\, F(k)\varphi_{1,k}(x)e^{iky},
\label{eq:Psi0}
\end{eqnarray}
where $F(k)$ is the Fourier transform of a 1D Gaussian function along the
$y$ axis:
\begin{eqnarray}
F(k) & = & \frac{1}{2\pi}\int dy\, 
e^{-iky}\frac{e^{-\frac{(y-y_0)^{2}}{4\sigma^{2}}}e^{+ik_{0}y}}
{\sqrt[4]{2\pi\sigma^{2}}}
\label{eq:ESWP}\\
&=& \sqrt[4]{\frac{\sigma^{2}}{2\pi^{3}}}e^{-\sigma^{2}(k-k_{0})^{2}}e^{-iky_{0}}.
\nonumber 
\end{eqnarray}
This leads to a localized WF along both directions:
in $y$, the Gaussian envelope gives a finite
extension around the central position $y_0$,
in $x$, the functions $\varphi_{1,k}(x)$ are
always localized around $x_{0}(k)$, and the wave vector $k$ is,
in turn, localized around $k_0$ by our choice of $F(k)$.
With the initial condition Eq.~(\ref{eq:Psi0}), 
our simulations will be able to take into account
the effects of the size $\sigma$ of the wave packet (WP) on the
interference phenomena.
In order to use Eq.~(\ref{eq:Psi0}), we will be careful to initialize
the WP $\Psi_0$ where the potential $V(x,y)$ does not depend on $y$.

\begin{figure}[h]
\begin{centering}
\includegraphics[scale=0.3]{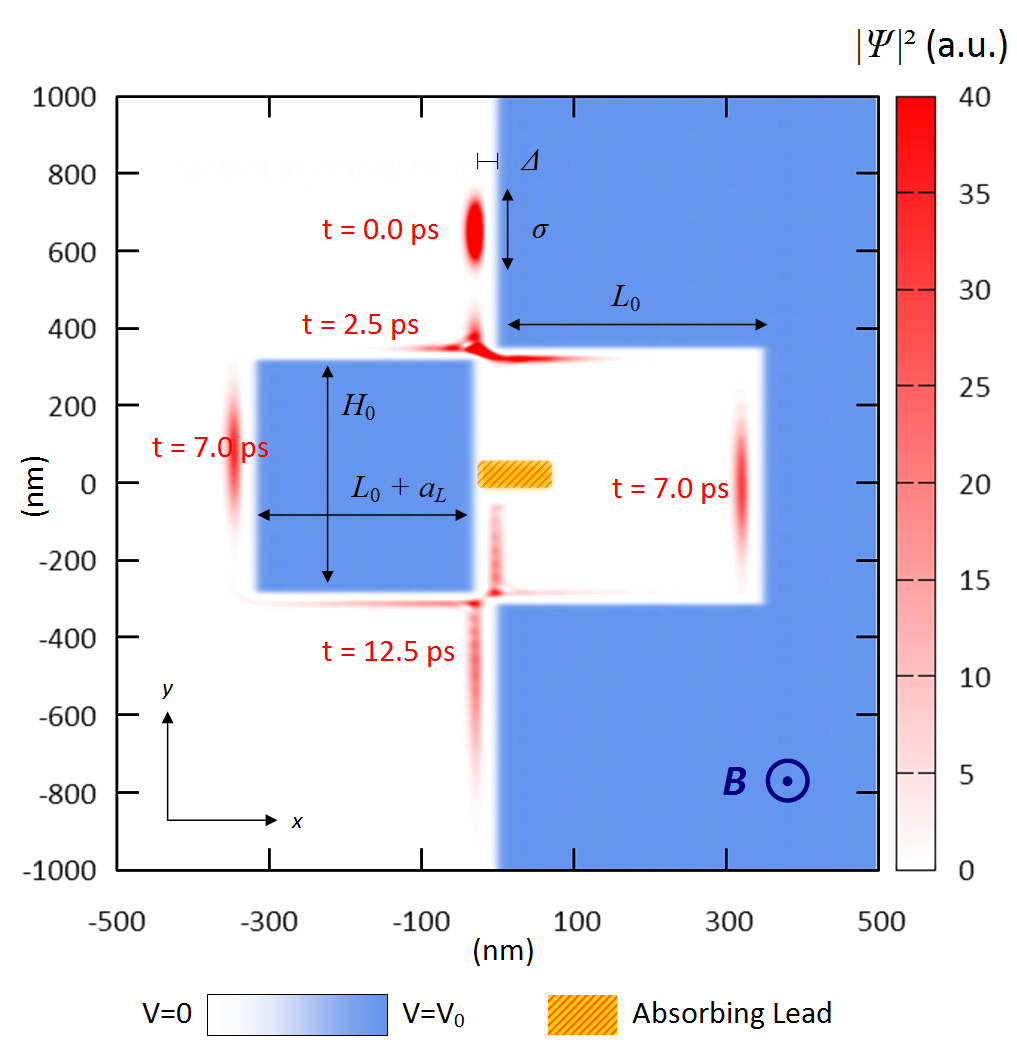}
\par\end{centering}
\caption{(Color online) Device geometry and square modulus of
the WF at $t=0$ (initial state, with $\sigma=40$~nm),
$t=2.5$~ps (at the first QPC),
$t=7$~ps (along the two arms),
$t=12.5$~ps (at the second QPC).
In this example $B=5$~T and $a_{L}=-65$~nm.
The resulting transmission is $T=0.599$.}
\label{fig:device}
\end{figure}

The specific form of the ESs depends on the the shape of the potential
barrier $V(x)$, and in general there is no closed-form expression
for them. 
In order to have a realistic model of the smooth edges between the allowed
2DEG region and the depleted one, defined e.g. with the split-gate technique,
we take 
\begin{equation}
V(x)=V_{0}\mathcal{F}_{\tau}(x),
\end{equation}
where $\mathcal{F}_{\tau}(x)=(\exp(\tau x)+1)^{-1}$ is a Fermi distribution
with a ``broadening'' parameter $\tau$,
and we compute numerically the corresponding $\varphi_{n,k}(x)$ states.

The shape of the depleted regions of our 2DEG is chosen in order to
mimic, with the ES channels, the Mach-Zehnder interferometer (MZI) of
Ref.~\onlinecite{JiMZI03}.
This is depicted in Fig.~\ref{fig:device}, where the dark and light regions
represent $V(x,y)=0$ and $V(x,y)=V_0$, respectively.
Two narrowings, where two areas with high potential are close to each other,
form two quantum point contacts (QPCs), both with a square area
\footnote{In the region of the QPCs
two potential barriers with a right-angle shape face each other. 
The prolongation of the edges of the angles identifies a square area
with side $\delta$. 
The position of the edges is taken at the inflection point of each barrier.
}  of $\delta^{2}$: their dimension $\delta$ is tuned
in order to give a  transmissivity of $50\%$ in the two output channels.
The localized electron is injected from the top of Fig.~\ref{fig:device}
(where the initial WF is centered around $x=-20$~nm and $y=700$~nm),
in the first LL, whose ESs follow the boundary between the high- and
low-potential regions.
After the first QPC, each ES is split into two parts 
that constitute the two arms (``left'' and ``right'') of the 
interferometer. They are rejoined at the second QPC.
As a consequence, the two components of the WP that
follow the two arms, interfere at the second QPC.
Indeed, two outputs are available there to the electron:
one part is reflected inside the MZI, and then absorbed by a 
contact (modeled as an absorbing imaginary potential $V_{abs}$,
as detailed later), the other part is transmitted towards the
bottom of the device.
The norm of the latter part gives a measure of the
transmission coefficient $T$ of the device.

The geometrical parameters of the MZI are
indicated in Fig.~\ref{fig:device}.
Specifically, the ``right'' arm and the ``left'' arm have two horizontal
sections each, of length $L_{0}$ and $L_{0}+a_{L}$, respectively.
The vertical sections of the two arms have obviously the same length, $H_0$.
The absorbing lead is modeled by a purely imaginary
potential\cite{kramer2008efficient,kramer2011time-dep}
\begin{equation}
V_{abs}(x,y) = iV_{abs}^{0} \frac{\mathcal{F}_{\tau}(-x+x_{c})
\mathcal{F}_{\tau}(x-x_{c}-\delta_{c}) }{
\cosh^{2}\left((y-y_{c})^{2}/d^{2}\right)},
\end{equation}
where $x_c$ and $y_c$ are the coordinates of the contact, while $\delta_c$ and $d$
are, respectively, the contact extension along the $x$ and $y$ axes.
By changing the parameters of the device, such as $a_{L}$
(which controls the relative length of the two arms and the
area) or the magnetic field $B$,
we observe AB oscillations
in the transmission coefficient $T$. 
With the introduction of the
absorbing lead $V_{abs}$, the calculation of $T$ is straightforward,
since it is the norm of the final wave function after the
absorption process. 
Differently to other works\cite{PhysRevB.90.161305_Gaury} focusing on the
dynamics of carriers moving along Hall ESs, our confining potential $V$
does not depend on time.

The ``center'' $x_{0}(k_{0})$ of the initial WP ($t=0$) is fixed at a
distance $\Delta$ from the inflection point of the potential barrier $V(x)$,
and consequently $k_{0}$ can be calculated by Eq.~(\ref{eq:k0-x0}). 
The resulting local bandstructure of the 1st LL around $k=k_{0}$ can be
determined numerically, and from a parabolic fit we get the values of the
parameters $E_0$, $m_{B}^{*}$ and $k_1$ in its expression:
\begin{equation}
E_{1}(k)=\frac{\hbar^{2}}{2m_{B}^{*}}(k-k_{1})^{2}+E_{0}.
\end{equation}
The parameters $k_{0}$ and $k_{1}$ are gauge-dependent quantities, 
since they depend by the origin of the coordinate system: however, 
with a proper choice of this degree of freedom and of the energy zero,
we can set $k_{1}=0$ and $E_{0}=0$ without any physical change
in the description of the system, in order to get a simpler description
of the system. 
Now $k_{0}$ is directly related to the group velocity
$v_{g}=\hbar k_{0}/m_{B}^{*}$ of the WP,
which behaves like a free particle of mass $m_{B}^{*}$ in 1D.

In the simulations we use for the electron the effective mass
of GaAs $m^*=0.067 \, m_e$.
The parameters of the potential and of the absorbing
contact are $V_{0}=10$~eV, $\tau^{-1}=3$~nm, $\delta_{c}=100$~nm,
$d=30$~nm, $V_{abs}^{0}=-98.7$~eV; the constructive parameters
of the device are $\delta=32.2$~nm, $L_{0}=350$~nm and $H_0=600$~nm;
the parameters of the WP are $\Delta=20$~nm and $\sigma=20,40,60$~nm%
\footnote{We choose these values for the standard deviation $\sigma$ of the
WPs because they result a good compromise for the simulations: indeed,
these WPs are small enough to be localized inside the device and at
the same time their time spreading is small enough to keep this condition
true during all the simulation time.%
}, while the magnetic field is kept around the value of $B=5$~T. \\

\section{Numerical simulations\label{sec:IIINumerical-simulations}}

The time evolution of the WP is realized with the split-step Fourier
method\cite{bertoni_serafini,kramer2008efficient,kramer2011time-dep}.
In summary, the evolution operator 
$\hat{U}(\delta t)=e^{-\frac{i}{\hbar}\delta t\cdot\hat{H}}$
is applied $N$ times to the initial wave function $\Psi(x,y;\,0)$,
each leading to the evolution of a short time step $\delta t$:
\begin{equation}
\Psi(x,y;\, t)|_{t=N\delta t}=[\hat{U}(\delta t)]^{N}\Psi(x,y;\,0) .
\label{eq:psiUevolution}
\end{equation}
The Hamiltonian of Eq.~(\ref{eq:2D_Hamilt}) can be written as
$\hat{H}=\hat{T}_{1}(x,p_{y})+\hat{T}_{2}(p_{x})+\hat{V}(x,y)$, with
\begin{eqnarray}
\hat{T}_{1}(x,p_{y})=-\frac{\hbar^{2}}{2m^*}\frac{\partial^{2}}{\partial y^{2}}
-i\hbar \frac{e B x}{m^*} \frac{\partial}{\partial y}
+\frac{e^2 B^2}{2m^*}x^2,\quad
\end{eqnarray}
\begin{eqnarray}
\hat{T}_{2}(p_{x}) = -\frac{\hbar^{2}}{2m^*}
\frac{\partial^{2}}{\partial x^{2}}. \quad
\end{eqnarray}
By applying the Trotter-Suzuky factorization and the split-step Fourier
method, the operator $\hat{U}(\delta t)$ can be approximated with
\begin{widetext}
\begin{equation}
[\hat{U}(\delta t)]^{N}=e^{+\frac{i}{\hbar}\delta t\cdot\frac{\hat{V}}{2}}\left[e^{-\frac{i}{\hbar}\delta t\cdot\left(\hat{V}+\hat{V}_{abs}\right)}\mathscr{F}_{y}^{-1}e^{-\frac{i}{\hbar}\delta t\cdot\hat{T}_{1}}\mathscr{F}_{y}\mathscr{F}_{x}^{-1}e^{-\frac{i}{\hbar}\delta t\cdot\hat{T}_{2}}\mathscr{F}_{x}\right]^{N}e^{-\frac{i}{\hbar}\delta t\cdot\frac{\hat{V}}{2}},
\label{eq:splitstepU}
\end{equation}
\end{widetext}
where $\mathscr{F}_{x(y)}$ and $\mathscr{F}_{x(y)}^{-1}$ denote, respectively,
the direct and inverse Fourier transform with respect to the variable $x(y)$. 
By using the above expression,
the numerical solution of Eq.~(\ref{eq:psiUevolution})
is reduced to an alternating application of discrete Fourier transforms and
array multiplications, since each operator acts only in its diagonal
representation.

Within this scheme, we can reproduce the exact dynamics
of the electronic WF inside the device. As an example,
the evolution of the WP at four different times
is shown in Fig.~\ref{fig:device} for the specific
case described in the caption.
The simulations have been performed with a time-step $\delta t=0.1$~fs.

\begin{figure}[b]
\begin{centering}
\includegraphics[scale=0.5]{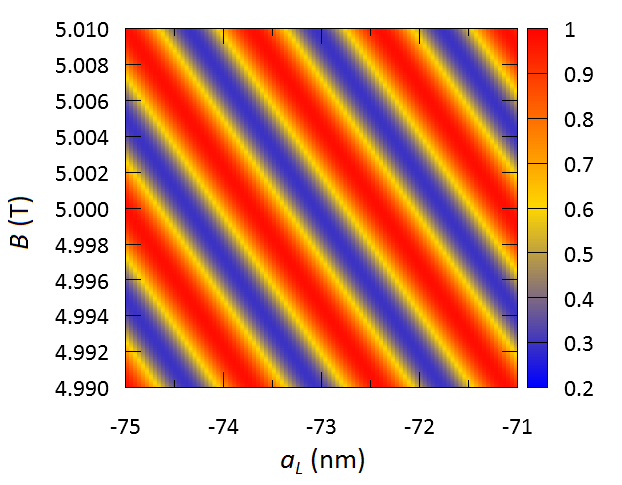}
\par\end{centering}

\caption{(Color online) Aharonov-Bohm oscillations in the transmission coefficient
$T$ as a function of the asymmetry parameter of the left arm $a_{L}$
and of the magnetic field $B$ (numerical simulations for a wave packet
with $\sigma=40\, \mathrm{nm}$).}
\label{fig:T_AB_oscill}
\end{figure}

As expected\cite{JiMZI03,Neder_DoubleMZI,Weisz_ERASER},
we obtain AB oscillations of the
transmission $T$ (see Figure \ref{fig:T_AB_oscill})
with respect to both the variation of the area (which is controlled
through $a_{L}$) and the variation of the magnetic field $B$.
This is consistent with the results of 
Refs.~\onlinecite{JiMZI03} and \onlinecite{Neder_DoubleMZI},
where the tailoring of the length of one of the two paths is achieved
through a modulation gate. 

However, we observe two extra features, namely (a) a damping of the AB
oscillations with $a_{L}$ and (b) an increase of the average $T$ as the
initial spatial localization $\sigma$ of the WP is increased,
with a consequent decrease of the visibility\cite{JiMZI03,Neder_PRL_MZILobes,
Neder_PRL_detectoreraser,Neder_DoubleMZI}
\begin{equation}
v_{MZI}=\frac{I_{max}-I_{min}}{I_{max}+I_{min}}=\frac{T_{max}-\langle T \rangle}{\langle T \rangle}
\end{equation}
of the AB oscillations (of the transmitted current $I$),
which is always smaller with respect to the ideal case $v_{MZI}=1$.
The two issues above are addressed in the following.

\begin{figure}
\begin{centering}
\includegraphics[scale=0.37]{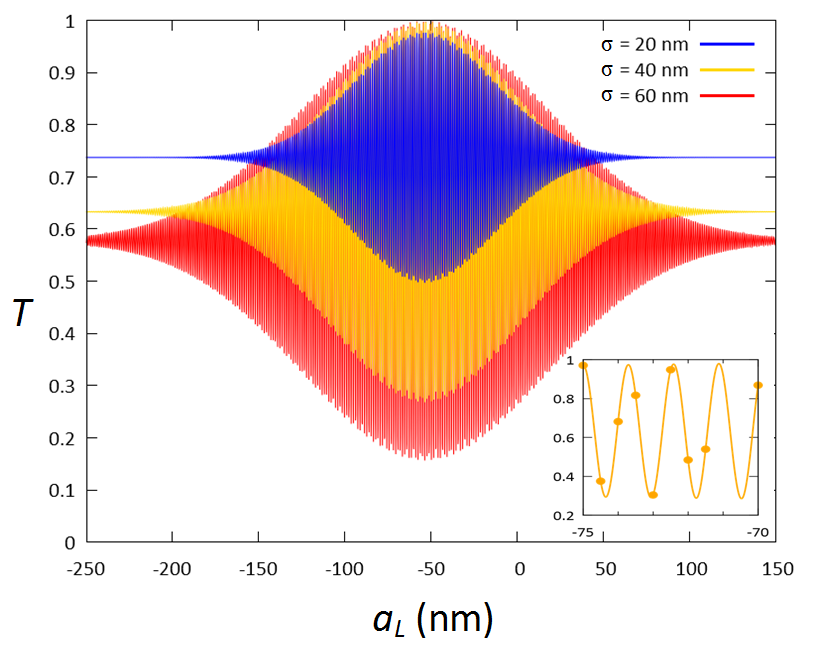}
\par\end{centering}

\caption{(Color online) Transmission coefficient $T$ as a function of the
asymmetry parameter of the left arm $a_{L}$
(top curve: $\sigma=20\, \mathrm{nm}$, middle curve: $\sigma=40\, \mathrm{nm}$, bottom curve:
$\sigma=60\, \mathrm{nm}$).
The curves are the fits of the numerical simulation data. 
It turns out that the best fit is obtained with the function 
$T(a_L)=c_1+c_2 \cdot\exp[-c_3(a_L-c_4)^2]\cdot\cos(c_5 a_L + c_6)$,
where the $c_i$ are the fit parameters.
Inset: Detail of the curve $\sigma=40\, \mathrm{nm}$
with numerical simulation data-points. All the simulations are performed
at $B=5$~T.}
\label{fig:T_al_oscill}
\end{figure}

(a) The AB oscillations
as a function of $a_{L}$ are modulated by a Gaussian envelope, whose
extension is directly correlated with the size $\sigma$ of the initial
WP. A fitting of $T$ vs $\sigma$ is reported in Fig.~\ref{fig:T_al_oscill},
for three values of $\sigma$ (see caption). 
This effect has a simple physical explanation,
already advanced by Haack et al. in \onlinecite{PhysRevB.84.081303_Haack}
for a Lorentzian WP. 
Indeed, when the asymmetry between the two
arms is large with respect to the size of the WP, the two parts
of the WF arrive at the second QPC at different times, and
do not interfere. 
In this case, each part is transmitted with a probability of 50\%,
ending up with a total transmission of $T\simeq0.5$.
In general, the larger the time offset at which the centers of the two WPs
reach the second QPC, the less effective is the quantum interference.
Therefore, we expect a saturation value of $\sim0.5$  for $T(a_{L})$,
and an oscillation amplitude of $\sim0.5$.
However, our numerical simulations
show a different trend, i.e. issue (b). 

(b) For smaller values of $\sigma$, the average 
(or saturation value with $a_{L}$) of the transmission $T$ is higher
and the amplitude of the oscillations is lower (Fig. \ref{fig:T_al_oscill}).
These effects are due to the energy-dependent
features of the scattering process at the QPCs,
which are included automatically 
in the numerical simulations
based on the direct solution of the Schr\"odinger equation.
In fact, contrary to a delocalized scattering-state model,
where the particle is represented by a single-energy state,
our WP is composed by different ESs, as given in Eq.~(\ref{eq:Psi0}),
each with a slightly different energy.
As a consequence, the two parts of the WP, transmitted and reflected
by the QPC, have different spectral weights, which depend on $\sigma$.
Only in the limit of $\sigma\to\infty$, the WP is split into two
identical parts that give an ideal $50\%$ splitting with $T=0.5$.
This effect should be detectable in noise spectra of two-particle
scattering, as proposed in Ref.~\cite{JPCM27.245302_Marian}.
In order to have a better physical insight and to
give a quantitative assessment of this effect, we will include
the energy-dependent transmissivity of the QPCs in the analytical model
presented in the following section.

\section{Theoretical Model for the Interference\label{sec:IVTheoretical-Model-for}}

Following \emph{Weisz et al}\cite{Weisz_ERASER}, we will represent a 
single ES as a 1D plane wave, and we will
describe the scattering process and the transport inside the MZI with
the \emph{scattering matrix} formalism\cite{Ferry09_Transport_Nano}.
Of course, the effects of the scattering on the WP can be determined
from the linear superposition of the scattering of single plane waves.

\begin{figure}
\begin{centering}
\includegraphics[scale=0.5]{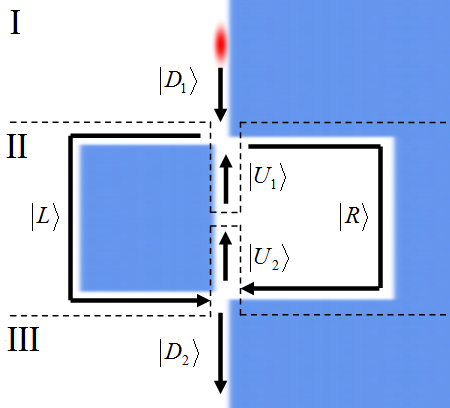}
\par\end{centering}

\caption{Scattering states inside the device. Regions I, II and III are
separated with dashed lines.}
\label{fig:scatt_states}
\end{figure}

The device can be divided into three regions – I, II and III –
as indicated in Fig.~\ref{fig:scatt_states},
where the plane-wave states are given by:
\begin{align}
I. &  & \left|D_{1}(k)\right\rangle =\left(\begin{array}{c}
1\\
0
\end{array}\right) &  & \left|U_{1}(k)\right\rangle =\left(\begin{array}{c}
0\\
1
\end{array}\right),\\
II. &  & \left|L(k)\right\rangle =\left(\begin{array}{c}
1\\
0
\end{array}\right) &  & \left|R(k)\right\rangle =\left(\begin{array}{c}
0\\
1
\end{array}\right),\\
III. &  & \left|D_{2}(k)\right\rangle =\left(\begin{array}{c}
1\\
0
\end{array}\right) &  & \left|U_{2}(k)\right\rangle =\left(\begin{array}{c}
0\\
1
\end{array}\right).
\end{align}

The initial wave packet is prepared in the state 
\begin{equation}
\left|\Psi_{I}\right\rangle =\int dk\, F(k)\left|D_{1}(k)\right\rangle ,
\end{equation}
 and the scatterings at the two QPCs are described by the
scattering matrices
\begin{equation}
S_{j}=\left(\begin{array}{cc}
r_{j} & i t_{j}\\
i t_{j} & r_{j}
\end{array}\right),
\end{equation}
with $j=1$ ($j=2$) labeling the first (second) QPC,
and where $r_{j}$ and $t_{j}$ are the transmission and reflection amplitudes.
After the first QPC, the WF is given by: 
\begin{align}
\left|\Psi_{II}\right\rangle  & =S_{1}\left|\Psi_{I}\right\rangle \nonumber \\
 & =\int dk\,\left(F_{L}(k)\left|L(k)\right\rangle +F_{R}(k)\left|R(k)\right\rangle \right),\label{eq:Psi_II}
\end{align}
where $F_{L}(k)=r_{1}(k)F(k)$ and $F_{R}(k)=i\, t_{1}(k)F(k)$. While
travelling in the two arms of the MZI inside region II, the WF acquires a different
phase in each arm, which is given by\cite{Ferry09_Transport_Nano,Heikkila_PNanoElect,KramerWGAB}:
\begin{equation}
\varphi_{h}=\frac{1}{\hbar}\int_{h}(\vec{p}-q\vec{A})\cdot ds
=\xi_{h}+\phi_{h},
\end{equation}
where $h=L$ ($h=R$) labels the left (right) arm and the 
integral is a line integral along the arm $h$.
The first contribution, the \emph{dynamic phase} $\xi_h$,
is due to the canonical momentum $\vec{p}$,
while the second contribution, the \emph{magnetic phase} $\phi_h$,
accounts for the vector potential $\vec{A}$.
The matrix describing the phase acquired is given by
\begin{equation}
P=\left(\begin{array}{cc}
e^{i\varphi_{L}} & 0\\
0 & e^{i\varphi_{R}}
\end{array}\right),
\end{equation}
and therefore the final WF in region III is given by
\begin{eqnarray}
\left|\Psi_{III}\right\rangle  
& = & S_{2}P\left|\Psi_{II}\right\rangle \nonumber \\
& = & \int dk\, F(k)\left(\tilde{t}\left|D_{2}(k)\right\rangle 
      +\tilde{r}\left|U_{2}(k)\right\rangle \right),
\end{eqnarray}
where $\tilde{t}=r_{1}r_{2}e^{i\varphi_{L}}-t_{1}t_{2}e^{i\varphi_{R}}$
and $\tilde{r}=ir_{1}t_{2}e^{i\varphi_{L}}-it_{1}r_{2}e^{i\varphi_{R}}$.
It is easy to see that the transmitted part of the WP is given by
\begin{align}
\left|\Psi_{III}^{T}\right\rangle =\left(\frac{1}{2\pi}\int dk'\left|D_{2}(k')\right\rangle \left\langle D_{2}(k')\right|\right)\left|\Psi_{III}\right\rangle ,\label{eq:Psi_III}
\end{align}
where $\left\langle D_{2}(k')|D_{2}(k)\right\rangle =2\pi\delta(k-k')$,
and therefore
\begin{eqnarray}
\tilde{T} & = & \left\langle \Psi_{III}^{T}\middle|\Psi_{III}^{T}\right\rangle ,\nonumber \\
 & = & 2\pi\int dk\,|F(k)|^{2}|\tilde{t}(k)|^{2}.\label{eq:T_numerical}
\end{eqnarray}
If the scattering amplitudes $r_{j}$ and $t_{j}$ are energy-independent,
Eq. (\ref{eq:T_numerical}) can be integrated analytically, to give:
\begin{equation}
\tilde{T}=T_{0}-T_{1}e^{-\frac{\Delta l^{2}}{8\sigma^{2}}}\cos(\Delta\phi-k_{0}\Delta l),\label{eq:T_model}
\end{equation}
where $T_{0}=\left|r_{1}r_{2}\right|^{2}+\left|t_{1}t_{2}\right|^{2}$
and $T_{1}=2r_{1}r_{2}t_{1}t_{2}$.
$T_{0}$ is the mean value of the oscillations,
while $T_{1}$ is their maximum amplitude. 
In the previous formula, $\Delta l\simeq2a_{L}$ is the
length difference between the two arms of the interferometer, while
$\Delta\phi\simeq\frac{e}{\hbar}BH_0(a_{L_0}+2L)=\pi\frac{\Phi_{B}}{\Phi_{0}}$
is the well-known Aharonov-Bohm phase\cite{Ferry09_Transport_Nano,jacoboni2010theory,Heikkila_PNanoElect}
($\Phi_{B}$ is the flux of the magnetic field through the area of
the MZI and $\Phi_{0}=\frac{h}{2e}$ is the \emph{magnetic flux quantum})
\footnote{The expressions for $\Delta\phi$ and $\Delta l$ should be corrected
with two geometrical parameters $c_{1}$ and $c_{2}$, which account
for the fact that the center of the ESs channels is inside the low-potential
region rather than on the allowed/forbidden region boundary, and
the area encircled is not a perfect rectangle.
Indeed, for $a_{L}=0$, the edge states travelling in the left arm follow
a longer path (external) with respect to those travelling in the right arm
(internal).  This is the reason why the curves obtained from numerical
simulations of Fig.~\ref{fig:T_al_oscill} are Gaussians which are not
centered around $a_{L}=0$.
We should write $\Delta l\simeq2a_{L}-2c_{1}$ and 
$\Delta\phi\simeq\frac{e}{\hbar}BH(a_{L}+2L-c_{2})=\pi\frac{\Phi_{B}}{\Phi_{0}}$
in order to take these effects into account.
However, this gives only an offset of the length difference of the arms
and does not affect the physics involved. For the sake of clarity we
do not include the above expressions in the analytical derivation.
}. 

\section{Discussion}
\label{sec:VDiscussion}

As one can see, Eq.~(\ref{eq:T_model}) provides the correct trend
for the results of the simulations, with a sinusoidal oscillation
enveloped by a Gaussian and with a fixed offset in the transmission.
Indeed, by substituting the expressions for $\Delta\phi$ and $\Delta l$  
into Eq.~(\ref{eq:T_model}), we see that the transmission coefficient
should oscillate as a function of $a_{L}$ like $\cos(k_{e}a_{L}+\varphi_{0})$,
where $k_{e}=\frac{e}{\hbar}BH-2k_{0}\simeq4.80\cdot10^{9}\, m^{-1}$,
which is consistent within 2\% with the values of $k_{e}$ obtained
by fitting the simulation data (see Table \ref{tab:fits}).
For what concerns the amplitude and the saturation value of $T$,
if we suppose that the QPCs have a perfect half-reflecting behavior
for all incoming energies – that is $t_{j}=r_{j}=\frac{1}{\sqrt{2}}$
for all $k$ – then our model predicts a saturation value $T_{0}=0.5$
and an oscillation amplitude $T_{1}=0.5$, as we previously stated.
This prediction, however, does not match with the numerical simulations,
where we observe higher values for $T_{0}$ and lower values for $T_{1}$
(see Table \ref{tab:fits}). 
The model also predicts that the oscillations of $T(a_{L})$ should be
modulated by a Gaussian with a standard deviation $\Sigma=\sigma$,
i.e. with the same spatial dispersion of the initial wave packet.
However, the actual values of $\Sigma$ obtained from numerical fitting
the data of Fig.~\ref{fig:T_al_oscill} 
are bigger than $\sigma$ by more than a factor of two in the case
with smaller $\sigma$. 
As $\sigma$ increases, the values of $\Sigma$ 
(always obtained by fitting the results of the simulations)
get closer to $\sigma$ (see Table \ref{tab:fits}).
Indeed, a better agreement can be obtained by enhancing our model.

This simplified model is valid under certain approximations,
that are: 
(i) the scattering process can be treated as a quasi-1D problem; 
(ii) all the edge channels involved in the transmission follow
the same path, i.e. we can consider the area enclosed by the interfering paths 
and the difference in length $\Delta l$ as they were independent on $k$;
(iii) the reflection and transmission amplitudes $r_{j}$ and $t_{j}$ 
of the QPCs are energy-independent (that means also $k$-independent).
While assumptions (i) and (ii) are in general verified, the approximation (iii) 
must be dropped in order to reproduce the results of the simulations.
Indeed, our results can be explained considering
that the scattering amplitudes of the QPCs are not energy-independent, as it has
been already pointed out in the literature \cite{Bocquillon_2Indisting_elec}. 
From a physical point of view, the edge states of higher energy
are closer to the barrier, and therefore they should tunnel easier
through the QPCs. This prediction can be easily verified, since the
energy-dependent scattering process should produce different values
in the weights $F_{L}(k)$ and $F_{R}(k)$ of the WP after the first
QPC. This can be linked directly to the differences in the values
of $T_{0}$ and $T_{1}$ with respect to the ideal case, as we will
see briefly (a model for the correct estimation of $\Sigma$,
$T_0$ and $T_1$ is given 
in Appendix \ref{sec:Appendix:-Calculation-of-Sigma}).

\begin{table}[H]
\begin{centering}
\begin{tabular}{|c|c|c|c|c|}
\hline 
\multicolumn{1}{|c|}{Initial WP} & \multicolumn{4}{c|}{$T(a_{L})$ (fit)}\tabularnewline
\hline 
$\sigma\;(\mathrm{nm})$ & $T_{0}$ & $T_{1}$ & $\Sigma\;(\mathrm{nm})$ & $k_{e}\;(\mathrm{nm^{-1}})$\tabularnewline
\hline 
\hline 
$20.0$ & 0.7371 & 0.2403 & $40.98$ & $4.86$\tabularnewline
\hline 
$40.0$ & 0.6330 & 0.3650 & $54.54$ & $4.86$\tabularnewline
\hline 
$60.0$ & 0.5778 & 0.4215 & $70.99$ & $4.86$\tabularnewline
\hline 
\end{tabular}
\par\end{centering}

\caption{Data of the initial WP ($\sigma$ is the standard deviation of Eq. (\ref{eq:ESWP}))
and the corresponding parameters for the transmission coefficient
$T(a_{L})$ profile. Data were fitted with the expression $T(a_{L})=T_{0}-T_{1}\exp(-\frac{1}{2}(a_{L}-c)^{2}/\Sigma^{2})\cos(k_{e}a_{L}+\varphi_{0})$, taken from Eq.~(\ref{eq:T_model}).}
 \label{tab:fits}
\end{table}

\begin{figure}
\begin{centering}
\includegraphics[scale=0.32]{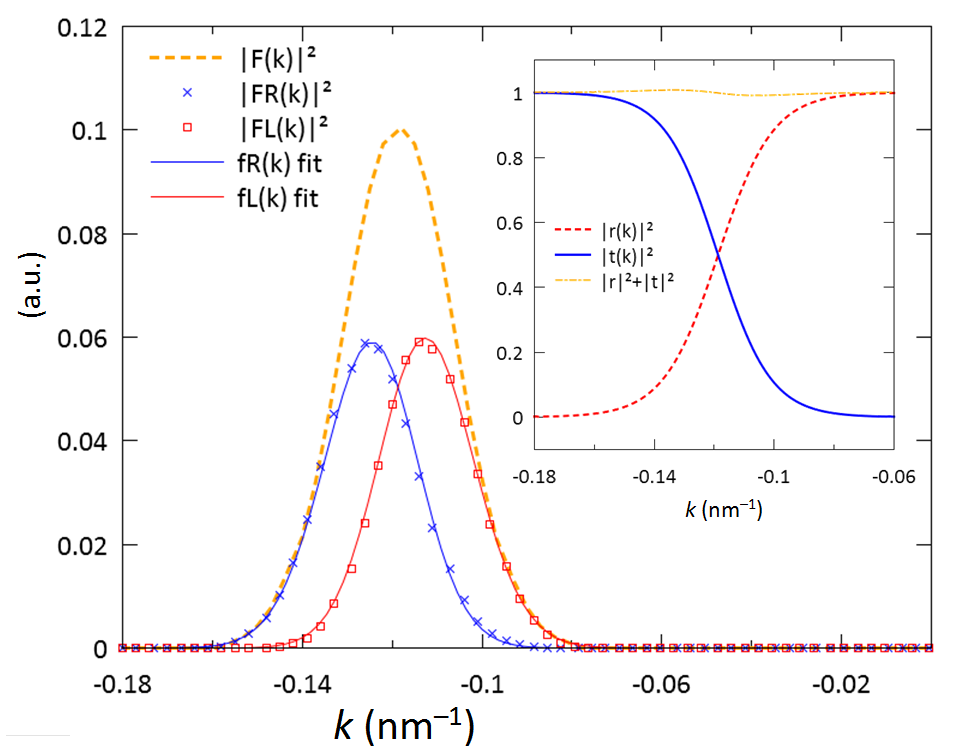}
\par\end{centering}

\caption{(Color online) Weights $|F_{R}(k)|^{2}$ and $|F_{L}(k)|^{2}$ for
the transmitted and reflected part of $\left|\Psi_{II}\right\rangle $
obtained from numerical simulations ($\sigma=40\, nm$); the continuous
lines are the fits with the functions $f_{R}(k)$ and $f_{L}(k)$,
the dashed line is $|F(k)|^{2}$ (which is peaked on $k=k_{0}$).
\textbf{Inset}: Squared modulus of transmission and reflection amplitudes
$t(k)$ and $r(k)$ for a single QPC calculated from numerical fits.}
\label{fig:T_r(k)_t(k)}
\end{figure}

The amplitudes $t(k)$ and $r(k)$ (which are equal for both QPCs,
due to their identical shape) can be calculated numerically from
$\left|\Psi_{II}\right\rangle $.
In Eq. (\ref{eq:Psi_II}), the squared modulus of the weights $|F_{L}(k)|^{2}$
and $|F_{R}(k)|^{2}$ has been calculated from a simulation on the
wave packet with $\sigma=40\, \mathrm{nm}$ by projecting $\left|\Psi_{II}\right\rangle $ over the local eigenstates 
$\left|L(k)\right\rangle$ and $\left|R(k)\right\rangle$ \cite{KramerWP}.
Then, using a phenomenological approach, the corresponding data have been 
fitted with the functions  $f_{L}(k)=\mathcal{F}_{\alpha_{r}}(k-k_{r})\,|F(k)|^{2}$
and $f_{R}(k)=\mathcal{F}_{\alpha_{t}}(k-k_{t})\,|F(k)|^{2}$, thus giving a 
Fermi-distribution expression\cite{buttiker1990quantized} for $|r(k)|^{2}$ and $|t(k)|^{2}$. The results of the fits
shown in Fig. \ref{fig:T_r(k)_t(k)} are in good agreement
with the numerical simulations and are also consistent with the constraint
$|r(k)|^{2}+|t(k)|^{2}=1$. From the same figure, we also deduce that
the QPCs are more transparent at higher energies (i.e., lower
values of $k$), as we predicted.

\begin{figure}
\begin{centering}
\includegraphics[scale=0.5]{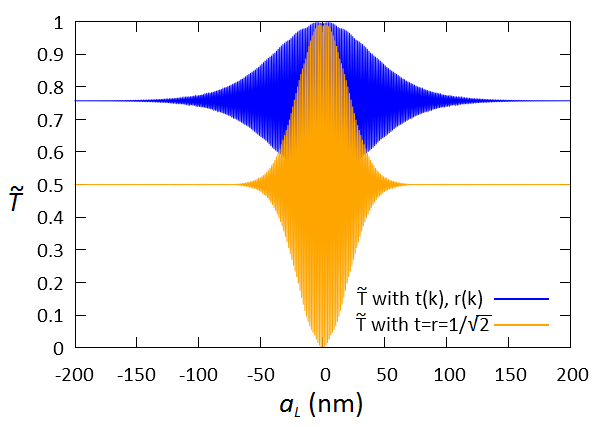}
\par\end{centering}

\caption{(Color online) Transmission coefficient $\tilde{T}$ calculated from the
theoretical model for $\sigma=20\, \mathrm{nm}$, when scattering amplitudes
$t=r=\frac{1}{\sqrt{2}}$ are energy-independent (Eq. (\ref{eq:T_model}),
bottom curve) and when scattering amplitudes are energy-dependent
(Eq.~(\ref{eq:T_numerical}), top curve). The energy-dependent curve
has $k_{e}=4.80\,\mathrm{nm}^{-1}$, $\tilde{T}_{0}=0.757$, 
$\tilde{T}_{1}=0.245$ and $\tilde{\Sigma}=40.4\,\mathrm{nm}$,
which are very close to the results of numerical simulations (see
Table \ref{tab:fits}). Note how the energy-dependent analytical 
model is able to reproduce the numerical results of 
Fig.~\ref{fig:T_al_oscill} (top curve)}
\label{fig:T_models_energy_dep}
\end{figure}

With the hypothesis that $t(k)$ and $r(k)$ are real\cite{Weisz_ERASER},
we can insert the transmission and reflection amplitudes calculated numerically 
into Eq. (\ref{eq:T_numerical}) and find the exact transmission coefficient
$\tilde{T}$. The results for a WP of $\sigma=20\, \mathrm{nm}$ are shown in 
Fig.~\ref{fig:T_models_energy_dep}: here, the simplified transmission
coefficient of Eq. (\ref{eq:T_model}) (where $t$ and $r$ are energy-independent
and both equal to $1/\sqrt{2}$) is compared with the exact energy-dependent
calculation of Eq. (\ref{eq:T_numerical}), which reproduces 
the result of numerical simulations (see Fig. \ref{fig:T_al_oscill}
for comparison). Then, as we see, the energy dependence of the scattering
amplitudes $t(k)$ and $r(k)$ is able to account for the values
of $T_{0}$ and $T_{1}$ and also for the broadening of $\Sigma$
(the region in which we observe interference) with respect to the
original spatial extension $\sigma$ of the WP. Therefore, taking
into account these corrections, it is still possible to use the simplified
model of Eq. (\ref{eq:T_model}) to describe the interference,
provided that one uses the energy-dependent values for $T_{0}$, $T_{1}$
and $\Sigma$, that we name $\tilde{T}_{0}$, $\tilde{T}_{1}$
and $\tilde{\Sigma}$. They can be calculated from our model as:
\begin{align}
\tilde{T}_{0} & =2\pi\int dk\,|F(k)|^{2}\left(\left|r_{1}r_{2}\right|^{2}+
\left|t_{1}t_{2}\right|^{2}\right), \label{eq:T0Tilde}\\
\tilde{T}_{1} & \simeq 2\pi\int dk\,|F(k)|^{2}2r_{1}r_{2}t_{1}t_{2}, \label{eq:T1Tilde}
\end{align}
(see Appendix \ref{sec:Appendix:-Calculation-of-Sigma} for $\tilde{\Sigma}$ and for the 
details of the derivation of  $\tilde{T}_0$ and $\tilde{T}_1$).
In summary, we obtained an analytical formula for the visibility
$v_{MZI}$ of the device, which is
\begin{equation}
v_{MZI}=\frac{\tilde{T}_{1}}{\tilde{T}_{0}}.
\end{equation}
Indeed, we see that the device has a reduced visibility with respect
to the expected ideal value $v_{MZI}=T_{1}/T_{0}=1$, due to phase-averaging
effects between different energies involved in the WP (and not because
of decoherence effects), as some experimental works already pointed
out\cite{JiMZI03}. Interestingly, these remarkable
effects, which are comparable with experimental observations, are
not explicitly related to fluctuations of the area enclosed by the
two paths of the interferometer, because we used the same values
of $\Delta l$ and $\Delta\phi$ for all edge states
composing the WP. 

As we can see from the model, in the mono-energetic plane-wave limit
($\sigma\rightarrow\infty$) the energy-dependent effects
become less important, and we recover the ideal behavior with 
$T_{0}=T_{1}=\frac{1}{2}$
(notice that $r(k_{0})=t(k_{0})\approx\frac{1}{\sqrt{2}}$) and unitary
visibility. The Gaussian modulation of the AB oscillations of the transmission
coefficient disappears, since $|F(k)|^2\rightarrow (2\pi)^{-1}\delta(k-k_{0})$: This
is consistent with scattering-states models used in the literature\cite{JiMZI03,
Neder_DoubleMZI,Weisz_ERASER}.

\section{Conclusions}
\label{sec:VConclusions}

In conclusion, our time-dependent simulations allow us to reproduce
the interference pattern of electrons transmitted through Landau
edge-states defining a MZI.
Specifically, we first used the split-step Fourier method to solve
exactly the time-dependent Schr{\"o}dinger equation,
and then we calculated the single-particle dynamics and the transmission
coefficient of the interferometer, which shows Aharonov-Bohm oscillations.
We exposed the effects of the finite size of the electronic WP.
In fact, the longitudinal extension (i.e. along the edge state) of the
WP is connected to its energy uncertainty which, in turn, is determined
by the injection process of the carriers: the larger the energy uncertainty,
the more localized the WP.
In Ref.~\onlinecite{JiMZI03}, for example, the carrier injection in the device
is achieved by an ohmic contact operating as the source terminal. 
However, the specific dynamics of the electrons moving from the Fermi sea
of the metallic contact to the semiconductor edge state is unknown, in general.
Indeed it is strongly affected by the atomistic details of the junction, by the
temperature (i.e. the broadening of the Fermi level) and, in the case
of Schottky contact, by the metal-semiconductor barrier.
Furthermore, if the injection of the single carrier is realized via
cyclic voltage pulses\cite{PhysRevLett.100.086601_Moskalets,Feve25052007,PhysRevB.90.235416_Hofer},
its WF can be partially tailored by the pulses timing and shape,
and the correlation of several carriers, possibly levitons, can be estimated
as reported in Ref.~\onlinecite{PhysRevB.91.195431}, thus assessing the transition
from single-particle to multi-particle regimes..
With our approach, that includes the energy selectivity of the QPCs, 
the real-space spreading of the carriers can be estimated
\emph{a posteriori} from the interference pattern\cite{PhysRevB.84.081303_Haack,PhysRevB.87.201302_Haack}.

In addition to simulating the exact dynamics,
we developed an analytical model that incorporates the physics
needed to reproduce the main features of the transport spectrum.
Such a model must go beyond the standard delocalized scattering-state
approach.
Indeed, we derived Eqs.~(\ref{eq:T_numerical}) and (\ref{eq:T_model}) by including
the effects of the finite size of the electronic WP.
Furthermore, the values of $T_0$ and $T_1$ in Eq. (\ref{eq:T_model}) must depend on
the different 
wave vectors $k$ composing the WP, in order to account for the energy-dependence
of the scattering at the QPCs and, finally, to reproduce the offset
value of the transmission, as a function of the spatial dispersion $\sigma$.
Also, the model leading to Eq.~(\ref{eq:T_numerical}) can justify
the reduction of the visibility of the interferometer
with respect to the ideal case. 

Finally, we stress that the understanding of the dynamics of single electrons
in Landau ESs, possibly including 4-way splittings in QPCs and a
non-trivial geometry, is of utmost importance for several proposals of
quantum computing architectures based on edge-channel transport.
Usually, in these proposals, the quantum bits are encoded into two ESs
of the same LL that are physically separated in space. However,
some recent studies also focused on the possibility of coupling
ESs belonging to different LLs\cite{venturelli2011edge_mixing,Giovannetti_Multichannel,Devyatov_copropagating_edge_interferometry},
between ESs belonging to two different 2DEGs through
tunneling\cite{Roddaro_Tunneling}
or between spin-resolved ESs of the same LL,
which could also be used in the future to realize a different type of
MZI\cite{karmakar2011controlled_coupling_SRES,karmakar2013towardsInterferom}
or scalable quantum gates.
Also in the above systems, the effects of the spatial localization of the WP
are pivotal for a realistic modeling of the devices.

\section*{Acknowledgements}

The simulations were performed at the facilities of CINECA within the
\emph{Iscra C} projects ``ESTENDI'' and ``TESSENA''. We would
like to thank Arianna Bergonzini, Carlo Maria Bertoni, Carlo Jacoboni 
and Enrico Piccinini for useful discussions.


\appendix
\section{Calculation of $\tilde{\Sigma}$, $\tilde{T}_0$ and $\tilde{T}_1$}
\label{sec:Appendix:-Calculation-of-Sigma}

In order to give an estimate for the energy-dependent value of $\tilde{\Sigma}$,
we will use a perfectly symmetrical form for $r(k)$ and $t(k)$, since
the QPCs are half-reflecting for $k=k_{0}$ and $r(k)^{2}+t(k)^{2}$
must be 1 for any $k$; furthermore, we do not distinguish between QPC 1 and 2:
\begin{eqnarray}
r(k) & = & \left[\exp(-\alpha(k-k_{0}))+1\right]^{-\frac{1}{2}} \\
t(k) & = & \left[\exp(+\alpha(k-k_{0}))+1\right]^{-\frac{1}{2}}
\end{eqnarray}
(where $\alpha=1.13\cdot10^{-7}m$ is obtained by fitting the numerical simulations).
We now replace these functions with Gaussian profiles, since near the turning point 
$k_{0}$ and in the decaying tail of the Fermi distribution it holds:
\begin{eqnarray}
\frac{1}{\exp(\alpha(k-k_{0}))+1}\simeq \exp\!\left[-\frac{(\alpha(k-k_{0})+e)^{2}}{4e}\right]\qquad
\end{eqnarray}
 (here $e$ is the Euler number). Therefore, since Eq.~(\ref{eq:T_numerical})
for $r_{1}=r_{2}=r$ and $t_{1}=t_{2}=t$
can be written as
\begin{eqnarray}
\tilde{T}= 2\pi\int dk\,|F(k)|^{2}\times \qquad \qquad \qquad \label{eq:Full_T_Appendix} \\
\times\left[r(k)^{4}+t(k)^{4}+2r(k)^{2}t(k)^{2}\cos(-k\Delta l+\Delta\phi)\right],\nonumber 
\end{eqnarray}
we deduce that, from each term in the square brackets, we can factor out
$\exp(-2\frac{\alpha^{2}}{4e}(k-k_{0})^{2})$, which can be multiplied
with $|F(k)|^{2}\propto\exp(-2\sigma^{2}(k-k_{0})^{2})$ to give an
``effective'' Gaussian weight with an increased standard deviation.
We conclude that the energy-dependent value for $\Sigma$
is given by:
\begin{equation}
\tilde{\Sigma}^{2}\simeq\sigma^{2}+\frac{\alpha^{2}}{4e}.
\end{equation}
With this approximated model we obtain, for $\sigma=20$~nm, $\tilde{\Sigma}=39.7$~nm,
which is in excellent agreement with the value $\Sigma=40.98$~nm obtained from the numerical
fitting of the exact simulations (see Table \ref{tab:fits}).
A similar good agreement is obtained for $\sigma=40$~nm and $\sigma=60$~nm.

Concerning the calculation of $\tilde{T}_0$ and $\tilde{T}_1$, we see from Eq. 
(\ref{eq:Full_T_Appendix}) that the oscillating part of $\tilde{T}$ and the Gaussian 
damping are given by the third term in the square brackets. 
Therefore, for large values of $\Delta l = 2 a_L$, the rapid oscillations of the cosine cancel out,
and only the first two terms survive, to give the saturation value of the transmission. Then
\begin{eqnarray}
\tilde{T}_0 & = & \underset{a_L \rightarrow \infty}{\lim}\tilde{T} \\
& = & 2\pi\int dk\,|F(k)|^{2} \left[r(k)^{4}+t(k)^{4}\right]. \nonumber
\end{eqnarray}
As we did for $\tilde{T}_0$, we can estimate $\tilde{T}_1$ by comparing Eq.~(\ref{eq:Full_T_Appendix})
with Eq.~(\ref{eq:T_model}).  
As a first approximation, $\tilde{T}_1$  is the maximum 
amplitude of the oscillating part, which is obtained when all the oscillating functions 
are maximized and the Gaussian damping is zero, i.e. when the argument $\Delta \varphi$
of all the cosines in the integral vanishes (and also $\Delta l=0$). 
Therefore, in this limit, we can subtract the background $\tilde{T}_0$ from $\tilde{T}$ 
to get $\tilde{T}_1$:
\begin{eqnarray}
\tilde{T}_1 & = & \underset{\Delta \varphi \rightarrow 0}{\lim}\tilde{T} - \tilde{T}_0\\
& = & 2\pi\int dk\,|F(k)|^{2} \left[2 r(k)^{2}t(k)^{2}\right]. \nonumber
\end{eqnarray}
For example, by using the above formulas, we obtain $\tilde{T}_0=0.760$ and
$\tilde{T}_1=0.240$ for the WP with $\sigma = 20$~nm, that are in good agreement
wit the values of  Table \ref{tab:fits}. 
It is straightforward to generalize these results to the cases when $r_1 \neq r_2$ and $t_1 \neq t_2$, to get
the expressions of Eqs. (\ref{eq:T0Tilde}) and (\ref{eq:T1Tilde}).

\bibliographystyle{apsrev4-1}

\bibliography{Paper_Edge_states_v11}

\clearpage
\onecolumngrid
\setcounter{figure}{0}

\section*{Supplementary data}

In this Supplementary text, we show, as an example, the dynamics of a
Lorentzian wave packet crossing a QPC of the kind described into
Sec.~\ref{sec:II-The-Physical-System} of the main text, and compare 
it with the corresponding result for a Gaussian wave packet, as
used in the simulations of our work.

We consider a 1D Lorentzian wave function
\begin{equation}
\psi_{1D}^L(y,t)= \frac{1}{\sqrt{\pi}}\frac{\sqrt{\frac{\Gamma}{2}}}
{\left(y-y_{0}\right)-v\,t+i\,\text{sgn}(k_{F})\frac{\Gamma}{2}} e^{ik_{F}(y+v\,t)},
\nonumber
\end{equation}
where $\Gamma$ is the full-width at middle height of the
probability density $|\psi_{1D}^L(y,t)|^{2}$, $y_0$ is the position of
the maximum at the initial time $t=0$.
For a system with a linear dispersion $E(k)=v\cdot k$, 
this corresponds to the wave function of a 
leviton\cite{PhysRevLett.97.116403_Keeling}, but it describes also an electron emitted
by a single-electron source quantum dot when subject to a suitable bias pulse
\cite{PhysRevLett.101.196404_Keeling}.
The imaginary exponential kinetic factor $\exp(ik_{F}(y+v\,t))$ provides
the group velocity $\langle v_{g}\rangle$ corresponding to the Fermi wave vector $k_{F}$, which is given below.
The factor $\text{sgn}(k_{F})$ is inserted
to select only the wavevectors $k$ associated
with an energy $E(k)>E_{F}$, as required by the leviton energy profile.

As in Eq.~(5) of the main text, we Fourier transform in the $k$ domain
in order to obtain proper weights for the ESs $\varphi(x,k)e^{iky}$
\begin{align} \label{eqFL_SUP}
F^L(k) &= \frac{1}{2\pi} \int dy\, e^{-iky}\psi_{1D}^L(y,0)
\tag{S1} \\ \nonumber
       &= -\sqrt{\frac{\Gamma}{\sqrt{2\pi}}} i e^{-\frac{1}{2}(k-k_{F})(2iy_{0}+\Gamma\text{sgn}(k_{F}))}
\Theta[(k-k_{F})\text{sgn}(k_{F})]\text{sgn}(k_{F}),
\end{align}
where we used the Heaviside function $\Theta[k]$. Therefore, the expectation value for the group velocity is:
\begin{equation}
\left\langle v_{g}\right\rangle  =\hbar\left\langle k\right\rangle 
=  k_{F}+\text{sgn}(k_{F})\frac{1}{\Gamma},
\nonumber
\end{equation}
from which we can define the central wavevector $k_0 = m \left\langle v_{g}\right\rangle / \hbar $ of the wavepacket. 
With the approximation that all the involved edge states have the same 
transverse profile (which we used also in Sect. \ref{sec:VDiscussion} 
to derive the analytical expression for the transmission coefficient), that is $\varphi(x,k)\approx\varphi(x,k_{F})$,
the 2D wavepacket becomes (see also Eq.~(4) of the main text)
\begin{align} \label{eqpsiL_SUP}
\psi^L(x,y) & = \int dk\, F^L(k)\varphi_{1,k}(x)e^{iky}  \tag{S2} \\ \nonumber
& \approx \varphi(x,k_{F}) \frac{e^{ik_{F}y}}{\sqrt{\pi}} 
\frac{\sqrt{\frac{\Gamma}{2}}}
{\left(y-y_{0}\right)+i\,\text{sgn}(k_F)\frac{\Gamma}{2}} ,
\end{align}
thus showing that our linear superposition of ESs has the required Lorentzian shape.
The above expression (S2) is taken as the initial Lorentzian wave packet
and is propagated with the numerical approach described in the main text
for the Gaussian case.
Supplementary Figures~1 and 2
report the square modulus of the Lorentzian and Gaussian wave functions,
respectively, at the initial time and at a following time step, after the action of the QPC. 
The wavepackets have been initialized to have the same group velocity 
(numerical parameters are given in the captions).
Furthermore, Supplementary Figure~3
shows the effect of the QPC energy selectivity on the Lorentzian
wave function.
Although the time evolution in the latter case is more noisy
than in the Gaussian case described in the main text, and the
spatial spreading is clearly larger, the QPC is able to split
the wave packet in two parts with essentially the same integrated
probability of $0.5$ but with different spectral components, as for
the Gaussian case described in the main text.

\begin{suppfigure}
\begin{centering}
\includegraphics[scale=0.4]{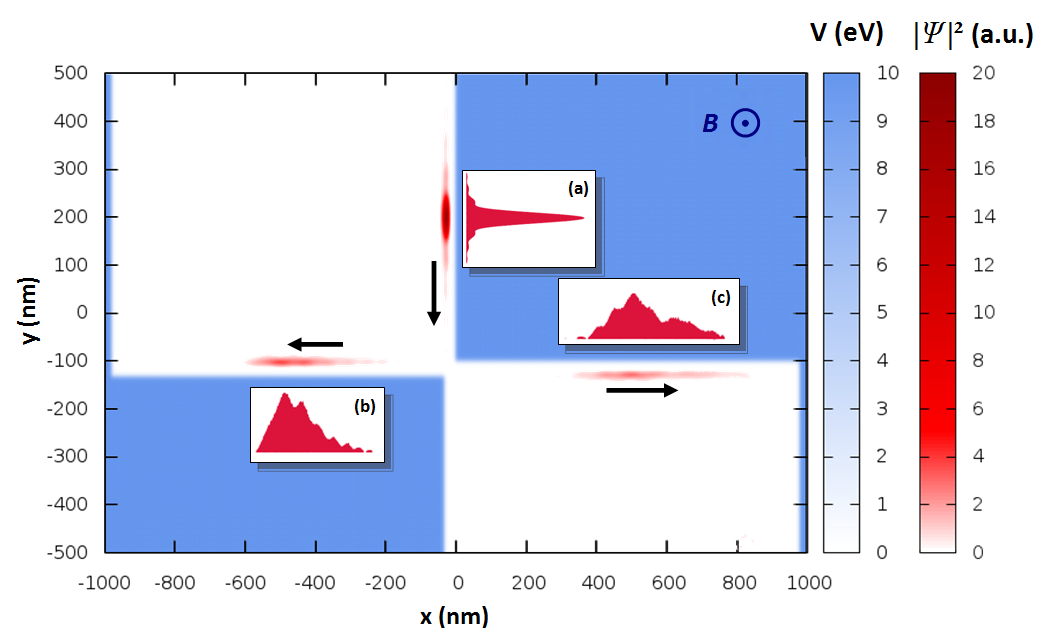}
\par\end{centering}
\caption{Square modulus of the Lorentzian wave function
of Eq.~(\ref{eqpsiL_SUP}) at the initial time (a) and at $t=6\,\text{ps}$, when
the WP has been split in a reflected (b) and a transmitted (c)
component by the QPC. Here $\Gamma=40\,\text{nm}$ and the wavevector 
$k_0$ associated with the group velocity is $\sim 1.2\,\text{nm}^{-1}$, while the other
parameters are the same as in Fig.~1 of the main text.}
\end{suppfigure}

\begin{suppfigure}
\begin{centering}
\includegraphics[scale=0.4]{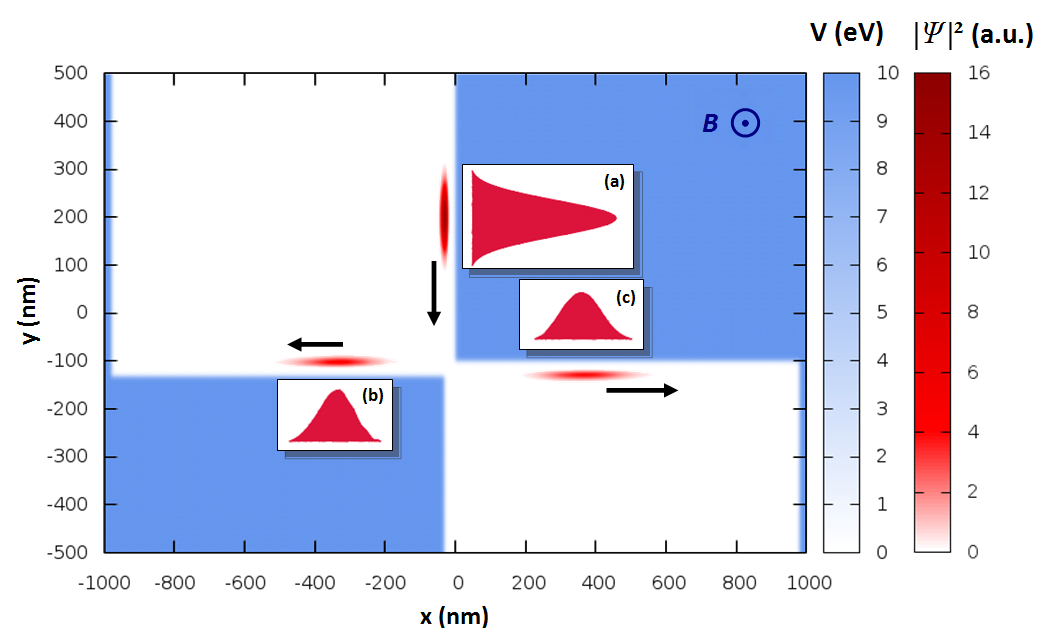}
\par\end{centering}
\caption{Square modulus of the Gaussian wave function
of Eq.~4 of the main text at the initial time (a) and at $t= 5\text{ps}$, when
the WP has been split in a reflected (b) and a transmitted (c)
component by the QPC. We used the same other
parameters as in Fig.~1 of the main text (i.e., $\sigma=40\,\text{nm}$ and
$k_0 \simeq 1.2\,\text{nm}^{-1}$).}
\end{suppfigure}

\begin{suppfigure}
\begin{centering}
\includegraphics[scale=0.3]{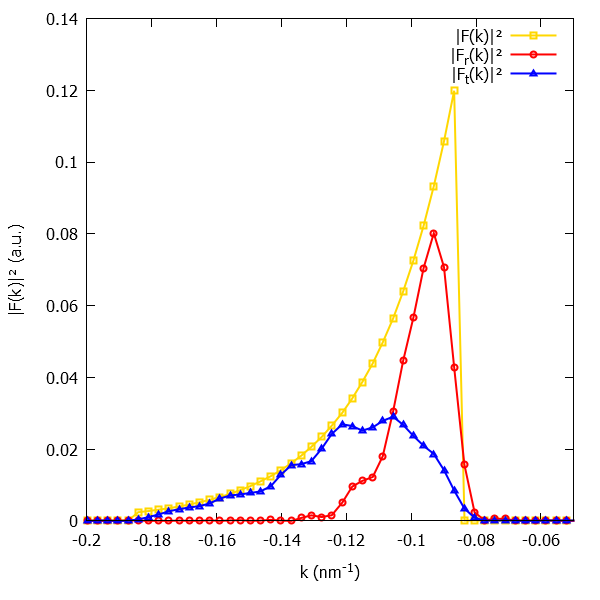}
\par\end{centering}
\caption{Reciprocal-space weights, as in Eq.~(\ref{eqFL_SUP}),
of the Lorentzian wave function at the initial time (yellow)
and for the transmitted (blue) and reflected (red)
components after the split at the QPC. Their real-space evolution is shown in
Supplementary Figure~1. }
\end{suppfigure}

\end{document}